\documentclass[a4paper,11pt]{article}
\date{}
\usepackage[latin1]{inputenc}
\usepackage[ruled,vlined,linesnumbered]{algorithm2e}
\usepackage{geometry}
\usepackage{amsmath,amsfonts,enumerate,amssymb,amsthm,color}
\usepackage{fullpage,enumerate}
\usepackage{graphicx}
\usepackage{latexsym,ifthen,epsfig,color}
\usepackage{float,paralist,cite}
\usepackage{version}

\newtheorem{Claim}{Claim}[section]
\newtheorem{theorem}{Theorem}
\newtheorem{lemma}{Lemma}

\newcommand{\join}{\text{\textcircled{{\footnotesize 1}}}}
\newcommand{\cojoin}{\text{\textcircled{{\footnotesize 0}}}}

\newcommand{\NP}{\ensuremath{\mathbb{NP}}}

\begin{document}

\author{
Andreas Brandst\"adt\\
\small Institut f\"ur Informatik, Universit\"at Rostock, D-18051 Rostock, Germany\\
\small \texttt{andreas.brandstaedt@uni-rostock.de}\\
\and
Raffaele Mosca\\
\small Dipartimento di Economia, Universit\'a degli Studi ``G. D'Annunzio'', 
Pescara 65121, Italy\\
\small \texttt{r.mosca@unich.it}
}

\title{On Chordal-$k$-Generalized Split Graphs}

\maketitle

\begin{abstract}
A graph $G$ is a {\em chordal-$k$-generalized split graph} if $G$ is chordal and there is a clique $Q$ in $G$ such that every connected component in $G[V \setminus Q]$ has at most $k$ vertices. Thus, chordal-$1$-generalized split graphs are exactly the split graphs. 

We characterize chordal-$k$-generalized split graphs by forbidden induced subgraphs. Moreover, we characterize a very special case of chordal-$2$-generalized split graphs for which the Efficient Domination problem is \NP-complete.    
\end{abstract}

\noindent{\small\textbf{Keywords}:
Chordal graphs;
split graphs;
partitions into a clique and connected components of bounded size;
forbidden induced subgraphs;
\NP-completeness; 
polynomial time recognition.
}

\section{Introduction}\label{sec:intro}

$G=(V,E)$ is a {\em split graph} if $V$ can be partitioned into a clique and an independent set. 
The famous class of split graphs was characterized by F\H{o}ldes and Hammer in \cite{FoeHam1977} as the class of $2K_2$-free chordal graphs, i.e., the class of $(2K_2,C_4,C_5)$-free graphs, 
and $G$ is a split graph if and only if $G$ and its complement graph $\overline{G}$ are chordal.

There are various important kinds of generalizations of split graphs: 
$G$ is {\em unipolar} if there is a clique $Q$ in $G$ such that $G[V \setminus Q]$ is the disjoint union of cliques of $G$, i.e., $G[V \setminus Q]$ is $P_3$-free.
Clearly, not every unipolar graph is chordal. $G$ is called a {\em generalized split graph} \cite{ProSte1992} if either $G$ or $\overline{G}$ is unipolar.  
Generalized split graphs are perfect, and Pr\"omel and Steger \cite{ProSte1992} showed that almost all perfect graphs are generalized split graphs.

We consider a different kind of generalizing split graphs: A graph $G$ is a {\em chordal-$k$-generalized split graph} if $G$ is chordal and there is a clique $Q$ in $G$ such that every connected component in $G[V \setminus Q]$ has at most $k$ vertices; we call such a clique $Q$ a {\em $k$-good clique of $G$}. Thus, chordal-$1$-generalized split graphs are exactly the split graphs. 
We characterize chordal-$k$-generalized split graphs by forbidden induced subgraphs.    

\medskip

An {\em induced matching} $M \subseteq E$ is a set of edges whose pairwise distance in $G$ is at least two. 
A {\em hereditary induced matching} ({\em h.i.m.}) is the induced subgraph of an induced matching, i.e., every connected component has at most two vertices, i.e., 
it is the disjoint union of an independent vertex set $S$ and the vertex set of an induced matching $M$ in $G$. 

Thus, $G$ is a chordal-$2$-generalized split graph if and only if $G$ has a clique $Q$ such that $G[V \setminus Q]$ is a hereditary induced matching.

\section{Basic Notions and Results}\label{sec:basic}

For a vertex $v \in V$, $N(v)=\{u \in V: uv \in E\}$ denotes its ({\em open}) {\em neighborhood}, and $N[v]=\{v\} \cup N(v)$ denotes its {\em closed neighborhood}. A vertex $v$ {\em sees} the vertices in $N(v)$ and {\em misses} all the others.
The {\em non-neighborhood} of a vertex $v$ is $\overline{N}(v):=V \setminus N[v]$.
For $U \subseteq V$, $N(U):= \bigcup_{u \in U} N(u) \setminus U$ and $\overline{N}(U):=V \setminus(U \cup N(U))$.   

\medskip

For a set ${\cal F}$ of graphs, a graph $G$ is called {\em ${\cal F}$-free} if $G$ contains no induced subgraph isomorphic to a member of ${\cal F}$.
In particular, we say that $G$ is {\em $H$-free} if $G$ is $\{H\}$-free.
Let $H_1+H_2$ denote the disjoint union of graphs $H_1$ and $H_2$, and for $k \ge 2$, let $kH$ denote the disjoint union of $k$ copies of $H$.
For $i \ge 1$, let $P_i$ denote the chordless path with $i$ vertices, and let $K_i$ denote the complete graph with $i$ vertices (clearly, $P_2=K_2$).
For $i \ge 4$, let $C_i$ denote the chordless cycle with $i$ vertices. A graph $G$ is {\em chordal} if it is $C_i$-free for any $i \ge 4$. 

\begin{figure}[ht]
  \begin{center}
   \epsfig{file=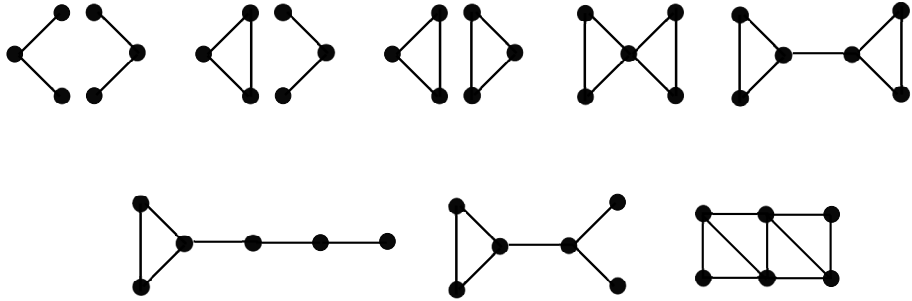}
   \caption{$2P_3$, $K_3+P_3$, $2K_3$, butterfly, extended butterfly, extended co-$P$, extended chair, and double-gem}
   \label{butterfly}
  \end{center}
\end{figure}

$H$ is a {\em butterfly} (see Figure \ref{butterfly}) if $H$ has five vertices, say $v_1,\ldots,v_5$ such that $v_1,v_2,v_3$ induce a $K_3$ and $v_3,v_4,v_5$ induce a $K_3$, and there is no other edge in $H$. 
$H$ is an {\em extended butterfly} (see Figure \ref{butterfly}) if $H$ has six vertices, say $v_1,\ldots,v_6$ such that $v_1,v_2,v_3$ induce a $K_3$ and $v_4,v_5,v_6$ induce a $K_3$, and the only other edge in $H$ is $v_3v_4$.

$H$ is an {\em extended co-$P$} (see Figure \ref{butterfly}) if $H$ has six vertices, say $v_1,\ldots,v_6$ such that $v_1,\ldots,v_5$ induce a $P_5$ and 
$v_6$ is only adjacent to $v_1$ and $v_2$. 

$H$ is a {\em chair} if $H$ has five vertices, say $v_1,\ldots,v_5$ such that $v_1,\ldots,v_4$ induce a $P_4$ and $v_5$ is only adjacent to $v_3$. 
$H$ is an {\em extended chair} (see Figure \ref{butterfly}) if $H$ has six vertices, say $v_1,\ldots,v_6$ such that $v_1,\ldots,v_5$ induce a
chair as above and $v_6$ is only adjacent to $v_1$ and $v_2$. 

$H$ is a {\em gem} if $H$ has five vertices, say $v_1,\ldots,v_5$ such that $v_1,\ldots,v_4$ induce a $P_4$ and $v_5$ is adjacent to $v_1,v_2,v_3$, and $v_4$.
  
$H$ is a {\em double-gem} (see Figure \ref{butterfly}) if $H$ has six vertices, say $v_1,\ldots,v_6$ such that $v_1,\ldots,v_5$ induce a gem, say with $P_4$ $(v_1,v_2,v_3,v_4)$ and $v_6$ is only adjacent to $v_3$ and $v_4$.  

\medskip

We say that for a vertex set $X\subseteq V$, a vertex $v \notin X$ has a join (resp.,~co-join) to $X$ if $X\subseteq N(v)$ (resp., $X\subseteq \overline{N}(v)$).
Join (resp., co-join) of $v$ to $X$ is denoted by $v \join X$ (resp., $v \cojoin X$); if $v \join X$ then $v$ is {\em universal for $X$}. Correspondingly, for vertex sets $X,Y \subseteq V$ with $X \cap Y = \emptyset$,
$X \join Y$ denotes $x \join Y$ for all $x \in X$ and $X \cojoin Y$ denotes $x \cojoin Y$ for all $x \in X$. A vertex $x \notin U$ {\em contacts $U$} if $x$ has a neighbor in $U$. For vertex sets $U,U'$ with $U \cap U' = \emptyset$, $U$ {\em contacts $U'$} if there is a vertex in $U$ contacting $U'$.   

\medskip

As a first step for our results, we show:

\begin{lemma}\label{existszjoinQ}
If $G=(V,E)$ is a chordal graph, $Q$ is a clique in $G$ and $Z$ is a connected component in $G[V \setminus Q]$ such that every vertex $q \in Q$ has a neighbor in $Z$ then there is a universal vertex $z \in Z$ for $Q$, i.e., $z \join Q$.
\end{lemma}

\noindent
{\bf Proof.}
We show inductively for $Q=\{q_1,\ldots,q_k\}$ that there is a $z \in Z$ with $z \join Q$: 

If $k=1$ then trivially, there is such a vertex $z \in Z$, and if $k=2$, say $Q=\{q_1,q_2\}$, then let $x,y \in Z$ be such that $q_1x \in E$ and $q_2y \in E$. If $xq_2 \notin E$ and $yq_1 \notin E$ then take a shortest path $P_{xy}$ between $x$ and $y$ in $Z$; clearly, since $G$ is chordal, there is a vertex in $P_{xy}$ seeing both $q_1$ and $q_2$. 
 
Now assume that the claim is true for clique $Q \setminus \{q_k\}$ with $k-1$ vertices and let $x \in Z$ be a universal vertex for $Q \setminus \{q_k\}$. 
If $xq_k \in E$, we are done. Thus assume that $xq_k \notin E$, and let $y \in Z$ with $q_ky \in E$. If $xy \in E$ then, since $G$ is chordal, clearly, $y$ is adjacent to all $q_i$, $1 \le i \le k-1$. Thus, assume that the distance between $x$ and $y$ in $Z$ is larger than 1. Without loss of generality, let $y$ be a neighbor of $q_k$ in $Z$ which is closest to $x$. Let $P_{xy}$ be a shortest path between $x$ and $y$ in $Z$. 
Thus, $q_k$ is nonadjacent to every internal vertex in $P_{xy}$ and to $x$. Then, since $G$ is chordal, $y$ is adjacent to all $q_i$, $1 \le i \le k-1$. 
Thus, Lemma \ref{existszjoinQ} is shown. 
\qed

\section{$k$-Good Cliques in Chordal Graphs}\label{goodcliques}

\subsection{$2K_2$-free chordal graphs}

For making this manuscript self-contained, let us repeat the characterization of split graphs (and a proof variant which can be generalized for $k$-good cliques). 
\begin{theorem}[\cite{FoeHam1977}]\label{splitgrchar}
A chordal graph $G$ is $2K_2$-free if and only if $G$ is a split graph. 
\end{theorem}

\noindent
{\bf Proof.} 
If $G$ is a split graph then clearly, $G$ is $2K_2$-free chordal.

\medskip

For the converse direction, assume to the contrary that for every clique $Q$ of $G$, there is a connected component, say $Z_Q$, of $G[V \setminus Q]$ with at least two vertices; we call such components {\em $2$-nontrivial}. Since $G$ is $2K_2$-free, all other components of $G[V \setminus Q]$ consist of isolated vertices.   

Let $Q_1:=\{q \in Q: q$ has a neighbor in $Z_Q\}$, and $Q_2 := Q \setminus Q_1$, i.e., $Q_2 \cojoin Z_Q$. Thus, $Q = Q_1 \cup Q_2$ is a partition of $Q$.
Since $G$ is $2K_2$-free and $Z_Q$ is 2-nontrivial, clearly, $|Q_2| \le 1$, and there is no connected component in $G[V \setminus (Q_1 \cup Z_Q)]$ with at least two vertices. Thus, $G[V \setminus (Q_1 \cup Z_Q)]$ is an independent vertex set.

Let $Q$ be a clique in $G$ with smallest 2-nontrivial component $Z_Q$ of $G[V \setminus Q]$ with respect to all cliques in $G$. Clearly, $Z_Q$ is also the nontrivial component of $G[V \setminus Q_1]$, i.e., $Z_{Q_1}=Z_Q$. Thus, $Q_1$ is a clique in $G$ with smallest 2-nontrivial component, and from now on, we can assume that every vertex in $Q_1$ has a neighbor in $Z_{Q_1}$. Thus, by Lemma \ref{existszjoinQ}, there is a universal vertex $z \in Z_{Q_1}$ for $Q_1$, i.e., $z \join Q_1$.

This implies that for the clique $Q':=Q_1 \cup \{z\}$, the 2-nontrivial component $Z_{Q'}$ (if there is any for $Q'$) is smaller than the one of $Q_1$ which is a contradiction. Thus, Theorem \ref{splitgrchar} is shown.
\qed  
 
\subsection{$(2P_3, 2K_3, P_3+K_3)$-free chordal graphs}

Clearly, a connected component with three vertices is either a $P_3$ or $K_3$, and graph $G$ is $(2P_3, 2K_3, P_3+K_3)$-free chordal if and only if $G$ is
$(2P_3, 2K_3, P_3+K_3,C_4,C_5,C_6,C_7)$-free. In a very similar way as for Theorem \ref{splitgrchar}, we show:

\begin{theorem}\label{3gensplit}
A chordal graph $G$ is $(2P_3, 2K_3, P_3+K_3)$-free if and only if $G$ is a chordal-$2$-generalized split graph.
\end{theorem}

\noindent
{\bf Proof.} 
If $G$ is a chordal-$2$-generalized split graph then clearly, $G$ is $(2P_3, 2K_3, P_3+K_3)$-free chordal.

\medskip

For the converse direction, assume to the contrary that for every clique $Q$ of $G$, there is a connected component, say $Z_Q$, of $G[V \setminus Q]$ with at least three vertices; we call such components {\em $3$-nontrivial}.  

Let $Q_1:=\{q \in Q: q$ has a neighbor in $Z_Q\}$, and $Q_2 := Q \setminus Q_1$, i.e., $Q_2 \cojoin Z_Q$. Thus, $Q = Q_1 \cup Q_2$ is a partition of $Q$.
Since $G$ is $(2P_3, 2K_3, P_3+K_3)$-free and $Z_Q$ is 3-nontrivial, clearly, $|Q_2| \le 2$, and there is no connected component in $G[V \setminus (Q_1 \cup Z_Q)]$ with at least three vertices. Thus, $G[V \setminus (Q_1 \cup Z_Q)]$ is a hereditary induced matching.

Let $Q$ be a clique in $G$ with smallest 3-nontrivial component $Z_Q$ of $G[V \setminus Q]$ with respect to all cliques in $G$. Clearly, $Z_Q$ is also the nontrivial component of $G[V \setminus Q_1]$, i.e., $Z_{Q_1}=Z_Q$. Thus, $Q_1$ is a clique in $G$ with smallest 3-nontrivial component, and from now on, we can assume that every vertex in $Q_1$ has a neighbor in $Z_{Q_1}$. Thus, by Lemma \ref{existszjoinQ}, there is a universal vertex $z \in Z_{Q_1}$ for $Q_1$, i.e., $z \join Q_1$.
This implies that for the clique $Q':=Q_1 \cup \{z\}$, the 3-nontrivial component $Z_{Q'}$ (if there is any for $Q'$) is smaller than the one of $Q_1$ which is a contradiction. Thus, Theorem \ref{3gensplit} is shown.
\qed  

\medskip

Clearly, not every $(2P_3, 2K_3, P_3+K_3)$-free graph and even not every $(2P_3$, $2K_3$, $P_3+K_3$, $C_5,C_6,C_7)$-free graph has a 3-good clique as the following example shows:

Let $v_1,v_2,v_3,v_4$ induce a $C_4$ with edges $v_iv_{i+1}$ (index arithmetic modulo 4) and let $x_1,x_2,x_3$ be personal neighbors of $v_1,v_2,v_3$ correspondingly.
Then any clique $Q$ has only at most two vertices, and for none of them, $G[V \setminus Q]$ is a hereditary induced matching. 

\subsection{The general case of $h$-good cliques in chordal graphs} 
 
As usual, for a pair of connected graphs $H_1$, $H_2$ with disjoint vertex sets, let $H_1+H_2$ denote the disjoint union of $H_1$ and $H_2$.  
For any natural $h$, let ${\cal C}_h$ denote the family of connected graphs with $h$ vertices, and let ${\cal A}_h = \{X + Y: X, Y \in {\cal C}_h\}$. In a very similar way as for Theorems \ref{splitgrchar} and \ref{3gensplit}, we can show:
 
\begin{theorem}\label{hgensplit}
For any natural $h$, chordal graph $G$ is ${\cal A}_{h+1}$-free if and only if there is a clique $Q$ of $G$ such that every connected component of $G[V \setminus Q]$ has at most $h$ vertices.
\end{theorem} 
 
\noindent
{\bf Proof.} 
If $G$ is a chordal graph with a clique $Q$ of $G$ such that every connected component of $G[V \setminus Q]$ has at most $h$ vertices, then clearly, $G$ is ${\cal A}_{h+1}$-free chordal.

\medskip

For the converse direction, assume to the contrary that for every clique $Q$ of $G$, there is a connected component,  say $Z_Q$, of $G[V \setminus Q]$ with at least $h+1$ vertices; we call such components {\em $(h+1)$-nontrivial}.  

Let $Q_1:=\{q \in Q: q$ has a neighbor in $Z_Q\}$, and $Q_2 := Q \setminus Q_1$, i.e., $Q_2 \cojoin Z_Q$. Thus, $Q = Q_1 \cup Q_2$ is a partition of $Q$.
Since $G$ is ${\cal A}_{h+1}$-free and $Z_Q$ is $(h+1)$-nontrivial, clearly, $|Q_2| \le h$, and there is no connected component in $G[V \setminus (Q_1 \cup Z_Q)]$ with at least $h+1$ vertices. Thus, every connected component of $G[V \setminus (Q_1 \cup Z_Q)]$ has at most $h$ vertices.

Let $Q$ be a clique in $G$ with smallest $(h+1)$-nontrivial component $Z_Q$ of $G[V \setminus Q]$ with respect to all cliques in $G$. Clearly, $Z_Q$ is also the nontrivial component of $G[V \setminus Q_1]$, i.e., $Z_{Q_1}=Z_Q$. Thus, $Q_1$ is a clique in $G$ with smallest $(h+1)$-nontrivial component, and from now on, we can assume that every vertex in $Q_1$ has a neighbor in $Z_{Q_1}$. Thus, by Lemma \ref{existszjoinQ}, there is a universal vertex $z \in Z_{Q_1}$ for $Q_1$, i.e., $z \join Q_1$.

This implies that for the clique $Q':=Q_1 \cup \{z\}$, the $(h+1)$-nontrivial component $Z_{Q'}$ (if there is any for $Q'$) is smaller than the one of $Q_1$ which is a contradiction. Thus, Theorem \ref{hgensplit} is shown.
\qed   
 
\medskip

Concerning the recognition problem, Tyshkevich and Chernyak \cite{TysChe1985} showed that unipolar graphs can be recognized in ${\cal O}(n^3)$ time. This time bound was slightly improved in \cite{EscWan2014}, and in \cite{McDYol2016}, McDiarmid and Yolov give a ${\cal O}(n^2)$ time recognition algorithm for unipolar graphs and generalized split graphs.
Clearly, for each fixed $k$, chordal-$k$-generalized split graphs can be recognized in ${\cal O}(n m)$ time since chordal graphs have at most $n$ maximal cliques,  and for each of them, say $Q$, it can be checked in linear time whether the connected components of $G[V \setminus Q]$ have at most $k$ vertices.
 
\section{Characterizing Split-Matching-Extended Graphs}\label{extsplitgr} 

$G=(V,E)$ is a {\em split-matching-extended graph} if $V$ can be partitioned into a clique $Q$, an independent vertex set $S_Q$ and the vertex set of an induced matching $M_Q$ in $G$ such that $S_Q \cojoin V(M_Q)$ and for every edge $xy \in M_Q$, at most one of $x,y$ has neighbors in $Q$ 
(note that $S_Q \cup V(M_Q)$ is a hereditary induced matching in $G$). Clearly, split-matching-extended graphs are chordal. Thus, split-matching-extended graphs are a 
special case of chordal-2-generalized split graphs. 

Clearly, split-matching-extended graphs can be recognized in linear time since for every edge $xy \in M$, the degree of $x$ or $y$ is 1, and deleting all such vertices of degree 1 leads to a split graph (which can be recognized in linear time). 

\medskip

Various algorithmic problems such as Hamiltonian Circuit and Minimum Domination are \NP-complete for split graphs. Efficient Domination, however, is solvable in polynomial time (even in linear time \cite{BraMos2016}) for split graphs while it is \NP-complete for split-matching-extended graphs (see e.g. \cite{BraMos2017}). 

\medskip

As mentioned in \cite{BraMos2017}, the graph $G_H$ in the reduction of Exact Cover 3XC to Efficient Domination is $(2P_3$, $K_3+P_3$, $2K_3$, butterfly, extended butterfly, extended co-$P$, extended chair, double-gem$)$-free chordal (and clearly, a special kind of unipolar graphs). Clearly, $2P_3$-free implies $C_k$-free for each $k \ge 8$. 

\begin{theorem}\label{extsplitgrchar}
$G$ is a split-matching-extended graph if and only if $G$ is $(C_4$, $C_5$, $C_6$, $C_7$, $2P_3$, $K_3+P_3$, $2K_3$, butterfly, extended butterfly, extended co-$P$, extended chair, double-gem$)$-free. 
\end{theorem}

\noindent
{\bf Proof.}
Clearly, if $G$ is a split-matching-extended graph then $G$ is $(C_4$, $C_5$, $C_6$, $C_7$, $2P_3$, $K_3+P_3$, $2K_3$, butterfly, extended butterfly, extended co-$P$, extended chair, double-gem$)$-free. 

\medskip

For the other direction, assume that $G$ is $(C_4,C_5,C_6,C_7,2P_3$, $K_3+P_3$, $2K_3$, butterfly, extended butterfly, extended co-$P$, extended chair, double-gem$)$-free. Then by Theorem \ref{3gensplit}, $G$ has a 2-good clique $Q$ such that $G[V \setminus Q]$ is a hereditary induced matching, say with independent vertex set $S_Q=\{s_1,\ldots,s_k\}$ and induced matching $M_Q=\{x_1y_1,\ldots,x_{\ell}y_{\ell}\}$. Let $Q$ be a maximum 2-good clique of $G$.

\begin{Claim}\label{N(Q)edgeneighbinclus}
For every edge $xy \in E$ with $x,y \in N(Q)$, $N_Q(x) \subseteq N_Q(y)$ and correspondingly for the non-neighborhoods, $\overline{N_Q}(y) \subseteq \overline{N_Q}(x)$ or vice versa, and in particular, $x$ and $y$ have a common neighbor and a common non-neighbor in $Q$.
\end{Claim}

\noindent
{\em Proof.} 
The claim obviously holds since $G$ is chordal and $Q$ is maximum.  
$\diamond$

\begin{Claim}\label{N(Q)edgeatleastonecommonneighbinQ}
For every edge $xy \in E$ with $x,y \in N(Q)$, $x$ and $y$ have exactly one common non-neighbor in $Q$. 
\end{Claim}

\noindent
{\em Proof.} Recall that by Claim \ref{N(Q)edgeneighbinclus}, $x$ and $y$ have a common neighbor, say $q_{xy} \in Q$, and 
suppose to the contrary that $x$ and $y$ have two common non-neighbors, say $q_1,q_2 \in Q$, $q_1 \neq q_2$. Then $x,y,q_{xy},q_1,q_2$ induce a butterfly in $G$ which is a contradiction. 
Thus, Claim~\ref{N(Q)edgeatleastonecommonneighbinQ} is shown.  
$\diamond$

\begin{Claim}\label{N(Q)edgeatleasttwoneighbinQ}
For every edge $xy \in E$ with $x,y \in N(Q)$, at least one of $x$ and $y$ has at least two non-neighbors in $Q$. Moreover, if $N_Q(x) \subseteq N_Q(y)$ then $y$ has exactly one non-neighbor in $Q$, and thus, $N_Q(x) \subset N_Q(y)$. 
\end{Claim}

\noindent
{\em Proof.} 
If both $x$ and $y$ would have exactly one non-neighbor, say $q_x,q_y \in Q$ with $xq_x \notin E$ and $yq_y \notin E$ then by Claim \ref{N(Q)edgeneighbinclus}, 
$q_x=q_y$ and now, $(Q \setminus \{q_x\}) \cup \{x,y\}$ would be a larger clique in $G$ but $Q$ is assumed to be a maximum clique in $G$. Thus, at least one of $x,y$ has at least two non-neighbors in $Q$. 

By Claim \ref{N(Q)edgeneighbinclus}, we have the corresponding inclusions of neighborhoods and non-neighborhoods in $Q$. 
If $N_Q(x) \subseteq N_Q(y)$ and $y$ has two non-neighbors in $Q$ then also $x$ has these non-neighbors but this contradicts 
Claim \ref{N(Q)edgeatleastonecommonneighbinQ}.
$\diamond$ 

\begin{Claim}\label{N(Q)2K2fr}
$G[N(Q)]$ is $2K_2$-free. 
\end{Claim}

\noindent
{\em Proof.} Suppose to the contrary that $G[N(Q)]$ contains $2K_2$, say $xy,x'y'$ with $xy \in E$ and $x'y' \in E$. Let $z \in Q$ be a common neighbor of $x,y$, and let $z' \in Q$ be a common neighbor of $x',y'$. If $z=z'$ then $x,y,x',y',z$ induce a butterfly. Thus, $z \neq z'$ for any common neighbors $z$ of $xy$, $z'$ of $x'y'$. If $x',y'$ miss $z$ and $x,y$ miss $z'$ then $x,y,x',y',z,z'$ induce an extended butterfly. If $x',y'$ miss $z$ but exactly one of $x,y$, say $x$, sees $z'$ then $x,z,x',y',z'$ induce a butterfly. If exactly one of $x,y$, say $x$, sees $z'$ and exactly one of $x',y'$, say $x'$, sees $z$ then  
$x,y,z,z',x',y'$ induce a double-gem. Thus, in any case we get a contradiction, and Claim \ref{N(Q)2K2fr} is shown.  
$\diamond$

\medskip

Thus, for at most one of the edges $x_iy_i \in M_Q$, both $x_i$ and $y_i$ have a neighbor in $Q$; without loss of generality, assume that $x_1$ and $y_1$ have a neighbor in $Q$, and for all other edges $x_jy_j \in M_Q$, $j \ge 2$, $y_j \cojoin Q$. By Claim \ref{N(Q)edgeneighbinclus}, let without loss of generality 
$N_Q(x_1) \subseteq N_Q(y_1)$, and let $u \in Q$ be a common neighbor of $x_1$ and $y_1$. 
By Claim \ref{N(Q)edgeatleasttwoneighbinQ}, $y_1$ has exactly one non-neighbor, say $z$, in $Q$ (which also misses $x_1$).   
Since $G$ is extended-co-$P$-free and butterfly-free, and by definition, $z \cojoin \{y_1,\ldots,y_{\ell}\}$, we have $z \cojoin \{x_1,\ldots,x_{\ell}\}$, 
and since $G$ is extended-chair-free and butterfly-free, $z$ has at most one neighbor in $S_Q$. 

Let $Q':= (Q \setminus \{z\}) \cup \{y_1\}$; clearly, $Q'$ is a clique in $G$. Then according to $Q'$, $G$ is a split-matching-extended graph, and thus, Theorem \ref{extsplitgrchar} is shown. 
\qed

\medskip

\noindent
{\bf Note added in proof.}
After writing this manuscript, we learnt that actually, the results of this manuscript are not new; they follow from previous papers of Gagarin \cite{Gagar1999} and of Zverovich \cite{Zvero2006}. Sorry for that! 

\begin{footnotesize}

\end{footnotesize}

\end{document}